\documentclass[10pt]{article}
\usepackage{graphicx,longtable,dcolumn,bm}
\begin{document}

\sloppy \emergencystretch=3pt
\begin{center}
{\Large \bf
 Study of Proton Induced Reactions in a Radioactive\\ $^{129}$I Target
at $E_p$=660 MeV}
\end{center}

\begin{center}
V. S. Pronskikh$^{1,}$\footnote[8]{Vitali.Pronskikh@jinr.ru},
J. Adam$^{1,2}$, A. R. Balabekyan$^{1,3}$, V. S. Barashenkov$^{1}$, 
V. P. Dzhelepov$^{1,\ \S}$, \\
 S. A. Gustov$^{1}$, 
V. P. Filinova$^{1}$, V. G.  Kalinnikov$^{1}$, M. I. Krivopustov$^{1}$, 
I. V. Mirokhin$^{1}$, 
A. A. Solnyshkin$^{1}$, \\
V. I. Stegailov$^{1}$, 
V. M. Tsoupko-Sitnikov$^{1}$, J. Mrazek$^{2}$, 
R. Brandt$^{4}$, W. Westmeier$^{4}$, R. Odoj$^{5}$, \\
S. G. Mashnik$^{6}$, A. J. Sierk$^{6}$, R. E. Prael$^{6}$, 
K. K.  Gudima$^{7}$, M. I. Baznat$^{7}$\\[5mm]
\end{center}

\begin{raggedright}
{\hskip 1 cm
1 Joint Institute for Nuclear Research, 141980 Dubna, Russia\\
\hskip 1 cm 2 Institute for Nuclear Physics, Academy of Sciences of the 
Czech Republic, \v{R}e\v{z}\\
\hskip 1 cm 3 Yerevan State University, Republic of Armenia\\
\hskip 1 cm 4 Institute of Nuclear Chemistry, Philipps University, 
Marburg, Germany\\
\hskip 1 cm 5 Forschungszentrum Julich, Germany\\
\hskip 1 cm 6 Los Alamos National Laboratory, Los Alamos, New Mexico 87545\\
\hskip 1 cm 7 Institute of Applied Physics, Academy of Science of Moldova, 
Chisinau\\
\hskip 1 cm \S\  Deceased}
\end{raggedright}

\vspace{0.3cm}
{\small \noindent
{\bf Abstract:}
Two NaI (85\% $^{129}$I and 15\% $^{127}$I) targets were 
exposed to a beam of 660-MeV protons.
Cross sections for formation of 76 residual 
nuclei were obtained by the induced activity method. The results are 
compared with other experimental data on $^{127}$I  and theoretical
calculations by eleven models contained in the codes \texttt{LAHET3} 
(using  the 
\texttt{Bertini+Dresner}, \texttt{ISABEL+Dresner}, \texttt{INCL+Dresner},
and \texttt{INCL+ABLA} options),
\texttt{CASCADE}, \texttt{CEM95}, \texttt{CEM2K}, \texttt{LAQGSM+GEM2},
\texttt{CEM2k+GEM2}, \texttt{LAQGSM+GEMINI}, and \texttt{CEM2k+GEMINI}.
Most of the models describe spallation products 
with masses close to the target reasonably well while the 
reliability of the codes differs greatly in the deep spallation 
and fission/fragmentation regions. 
The difficulties in describing products with A=40-80 by all of the 
codes tested here except for \texttt{CEM2k+GEMINI} and \texttt{LAQGSM+GEMINI} 
is related to the neglect of fission (and fragmentation) processes for 
targets as light as $^{129}$I.
}

\vspace*{5mm}
{\noindent \bf \large Introduction}\\

This work  is a continuation of~\cite{adam1}, where results  of
determining  cross sections for the yields of residual nuclei  from
the  $^{237}$Np and $^{241}$Am targets exposed to an extracted 660-MeV proton
beam of the JINR DNLP Phasotron were reported.
The  first  experiments to study transmutation of  $^{129}$I ($T_{1/2}
= 1.57\times 10^7$ y) and some transuranic isotopes were carried out in 1996
with  heavy-ion  beams  of the JINR LHE Synchrophasotron
using relativistic protons of 3.7 GeV~\cite{krivo2}.
Preliminary results of our measured product cross sections
from $^{129}$I targets irradiated with 660 MeV protons were published
earlier \cite{tsukuba}.
Here, we present final experimental cross sections for 
our 85\% $^{129}$I and 15\% $^{127}$I target
irradiated by 660 MeV protons.
Using previous
measurements on $^{127}$I targets generally performed 
at energies below 660 MeV analyzed by Molodo and Holzbach
\cite{modolo} and reducing them to our proton energy of 660 MeV by
linear interpolation between energies 600 and 800 MeV,
we estimate here experimental
cross sections for the target $^{129}$I at 660 MeV. We 
analyze our measurements with eleven models incorporated 
into several transport codes used recently in different applications.
Details on our measurement and on models used here to analyze the 
data may be found in Ref. \cite{PRC04}

\vspace*{5mm}
{\noindent \bf \large Results and Discussion}\\

Table~\ref{table1} presents our data: Column 1 lists
the isotopes (ground or metastable states) for which we obtained
reliable production cross sections, columns 2  and 3 show
their half-lives $(T_{1/2})$, decay and cross section types,
respectively,
column 4 presents the measured cross
sections for our  85\% $^{129}$I and 15\% $^{127}$I target,
column 5 shows cross sections for $^{127}$I from \cite{modolo},
and column 6 shows our deduced experimental cross sections
for $^{129}$I.

\begin{longtable}{|r|r|l|c|c|c|c|}
\caption{Experimental results}\label{table1}\\
\hline \hline
Residual&$T_{1/2}$&Decay&$\sigma_{exp}$,&$\sigma_{exp}$\cite{modolo}&$\sigma_{exp}$\\
&           &                 &    mb       &($^{127}$I), &($^{129}$I),\\
&           &                 &             &   mb        &  mb      \\  \hline
\endfirsthead
\caption{Experimental results (continued)}\\
\hline
\endhead
\hline
\endfoot
$^{44m}$Sc&    2.44 d&   I(IT,EC)          &   0.20(4)&     -----  &     ----- \\
$^{46}$Sc &   83.83 d&   C($\beta-$)       &   0.36(4)&     -----  &     ----- \\
$^{48}$V  &   15.97 d&   C($\beta+$,EC)    &   0.58(6)&     -----  &     ----- \\
$^{52}$Mn &    5.29 d&   C(EC,$\beta+$)    &   0.37(5)&     -----  &     ----- \\
$^{56}$Co &   78.8  d&   C(EC,$\beta+$)    &   0.11(4)&     -----  &     ----- \\
$^{58}$Co &   70.92 d&   I(EC,$\beta+$)    &   0.47(15)&    0.015(5)&    0.55(20) \\
$^{59}$Fe &   44.5  d&   C($\beta-$)       &   0.065(7)&    -----  &     ----- \\
$^{65}$Zn &  244.1  d&   C(EC,$\beta+$)    &   0.88(9)&     -----  &     ----- \\
$^{72}$As &   26    h&   I($\beta+$,EC)    &   0.93(9)&     -----  &     ----- \\
$^{72}$Se &    8.4  d&   C(EC)             &   0.59(15)&    -----  &     ----- \\
$^{74}$As &   17.78 d&   I(EC,$\beta+$,$\beta-$) &   0.85(9)&     -----  &     ----- \\
$^{76}$Br &   16.2  h&   C($\beta+$,EC)    &   0.77(9)&     -----  &     ----- \\
$^{77}$Br &    2.38 d&   C(EC,$\beta+$)    &   0.64(15)&    -----  &     ----- \\
$^{83}$Rb &   86.2  d&   C(EC)             &   0.40(8)&    0.53(5) &    0.38(10) \\
$^{84}$Rb &   32.87 d&   I(EC,$\beta+$,$\beta-$) &   0.12(4)&     -----  &     ----- \\
$^{85}$Sr &   64.84 d&   C(EC)             &   1.49(16)&    -----   &    ----- \\
$^{86}$Y  &   14.74 h&   C($\beta+$,EC)    &   0.69(25)&   1.04(10)&    0.63(30) \\
$^{87}$Y  &    3.35 d&   C(EC,$\beta+$)    &   1.15(11)&   1.35(15)&    1.11(15) \\
$^{88}$Y  &  106.6  d&   I(EC,$\beta+$)    &   0.36(10)&    2.21(25)&   ----- \\
$^{88}$Zr &   83.4  d&   C(EC)             &   1.4(4)&     -----  &     ----- \\
$^{89}$Zr &    3.27 d&   C(EC,$\beta+$)    &   1.51(15)&    1.54(30)&    1.50(17) \\
$^{90}$Nb &   14.60 h&   C($\beta+$,EC)    &   1.18(13)&    1.93(20)&    1.05(13) \\
$^{92m}$Nb&   10.15 d&   I(EC,$\beta+$)    &   0.11(4)&     -----  &     ----- \\
$^{93m}$Mo&    6.85 h&   I(IT,EC)    &   0.94(30)&   1.00(20)&    0.93(35) \\
$^{93}$Tc &    2.75 h&   C(EC,$\beta+$)    &   2.14(24) &     -----  &    ----- \\
$^{94}$Tc &    4.88 h&   I(EC,$\beta+$)    &   1.79(17)&    2.01(35)&    1.74(12) \\
$^{94m}$Tc&     52  m&   I($\beta+$,EC)    &   0.44(8)&     -----  &    ----- \\
$^{95}$Nb &   34.98 d&   C($\beta-$)       &   0.36(5)&    0.14(2) &    0.40(6) \\
$^{95}$Tc &   20.0  h&   C(EC)       &   3.00(34)&   5.2(7)&    2.6(4) \\
$^{96}$Tc &    4.28 d&   I(EC)       &   2.50(25)&   2.57(50)&    2.49(30) \\
$^{99}$Rh &   16.0  d&   I(EC,$\beta+$)    &   0.84(25)&   0.95(10)&    0.82(27) \\
$^{100}$Rh &   20.8  h&   I(EC,$\beta+$)    &   4.21(50)&   4.81(60)&    4.10(50) \\
$^{100}$Pd &    3.63 d&   C(EC)       &   2.85(33)&   4.1(4)&    2.6(4) \\
$^{101m}$Rh &   4.34 d&   C(EC,$\beta+$)    &   9.21(34)&   10.2(10)&    9.03(40) \\
$^{101m}$Rh&    4.34 d&   I(EC,$\beta+$)    &   2.9(9)&  ----- &   ----- \\
$^{101}$Pd &    8.47 h&   C(EC,$\beta+$)    &   6.3(8) &    6.9(7)  &    6.2(10) \\
$^{102m}$Rh&    2.9  Y&   I(EC)       &   2.98(30)&   2.34(25)&    3.09(35) \\
$^{103}$Ru &   39.25 d&   C($\beta-$)       &   0.43(5)&    0.30(3) &    0.45(6) \\
$^{104}$Ag &   69.2  m&   I(EC,$\beta+$)    &   8.3(8)&     ----- &     ------ \\
$^{105}$Ag &   41.29 d&   C(EC)       &  14.3(17) &  18.3(20) &   13.6(20) \\
$^{106}$Ag &    8.46 d&   I(EC)       &  7.5(7) &     6.9(8)&    7.6(8)\\
$^{108}$In &      58 m&   I($\beta+$,EC)    &   7.1(7)&     ----- &     ------ \\
$^{109}$In &    4.20 h&   C(EC,$\beta+$)    &  15.1(20)  &  10.7(12)  &  15.9(30)  \\
$^{109}$In &    4.20 h&   I(EC,$\beta+$)    &  12.1(12)  &  -----  &   -----  \\
$^{109}$Sn &   18.0  m&   C(EC.$\beta+$)    &   3.02(30)&     ----- &     ----- \\
$^{110m}$Ag&  249.9  d&   I($\beta-$,IT)    &   1.50(18)&   0.89(8) &    1.61(15) \\
$^{110}$In &    4.9  h&   I(EC)       &  11.7(12)  &  19.6(20) &   10.3(12) \\
$^{110m}$In&   69.1  m&   C(EC)       &   6.4(8)  &     ----- &     ----- \\
$^{113}$Sn &  115.1  d&   C(EC)       &  27.2(30) &  30.7(30) &   26.6(35) \\
$^{114m}$In&   49.51 d&   I(IT,EC)    &   7.1(7)&   4.8(5)&    7.5(7) \\
$^{114}$Sb &   3.49  m&   C(EC,$\beta+$)    &   2.92(35)&     ----- &     ----- \\
$^{115m}$In&    4.49 d&   I(IT,$\beta-$)    &  15.6(35) &     ----- &     ----- \\
$^{115}$Sb &   32.1  m&   C(EC,$\beta+$)    &  20.0(30) &     ----- &     ----- \\
$^{116m}$In&   54.15 m&   I($\beta-$)       &   2.72(42)&     ----- &     ----- \\
$^{116}$Sb &   15.8  m&   I($\beta+$,EC)    &   2.0(4)  &     ----- &     ----- \\
$^{116m}$Sb&   60.3  m&   I(EC,$\beta+$)    &  11.6(14) &     ----- &     ----- \\
$^{116}$Te &    2.49 h&   C(EC,$\beta+$)    &   9.9(10) &     ----- &     ----- \\
$^{117}$In &   43.8  m&   I($\beta-$)       &   1.6(3)  &     ----- &     ----- \\
$^{117}$Te &    1.03 h&   C(EC,$\beta+$)    &  15.4(15) &     ----- &     ----- \\
$^{118m}$Sb&    5.00 h&   I(EC,$\beta+$)    &  11.1(12)  &   8.8(12) &   11.5(17) \\
$^{118}$Te &    6.00 d&   I(EC)             &  13.9(14) &     ----- &     ----- \\
$^{118}$I  &   13.7  m&   C($\beta+$,EC)    &   3.3(4)  &     ----- &     ----- \\
$^{119}$Te &   16.05 h&   C(EC,$\beta+$)    &  11.5(12)  &  15.8(16) &   10.7(15) \\
$^{119m}$Te&    4.69 d&   I(EC,$\beta+$)    &  16.1(15)  &  15.5(20) &   16.3(20) \\
$^{120m}$Sb&    5.76 d&   I(EC)             &   6.3(6)&   5.6(6)  &    6.4(7) \\
$^{120}$I  &    1.35 h&   C($\beta+$,EC)    &  10.2(12) &     ----- &     ----- \\
$^{120m}$I &   53.0  m&   I($\beta+$,EC)    &   2.9(3)&     ----- &     ----- \\
$^{121}$Te &   16.8  d&   C(EC)       &  17.9(18)  &  24.0(40) &   16.8(20) \\
$^{121m}$Te&  154.0  d&   I(IT,EC)    &  13.3(16) &  16.9(17) &   12.7(13) \\
$^{122}$Sb &    2.70 d&   C($\beta-$,EC,$\beta+$) &   7.7(9)  &   4.3(5)&    8.3(10) \\
$^{123}$I  &   13.2  h&   C(EC)       &  25.0(24)  &  26.1(30) &   24.6(30) \\
$^{124}$I  &    4.18 d&   I(EC,$\beta+$)    &  26.3(30)  &  32.3(40) &   25.3(35) \\
$^{126}$Sb &   12.4  d&   I($\beta-$)       &   0.54(5) &     ----- &     ----- \\
$^{126}$I  &   13.02 d&   I(IT)       &  35(5)  &  66.0(80) &   29.5(20) \\
$^{127}$Xe &   36.46 d&   C(EC)       &   3.8(4)&   1.42(20)&    4.2(5) \\
$^{128}$I  &   25.0  m&   I($\beta-$,EC)    &  31(5) & ----- &  ----- \\\hline \hline
\end{longtable}

We analyzed our measurements with the LAHET3 version \cite{LAHET3}
of the transport code LAHET \cite{LAHET}
using the Bertini \cite{Bertini} and ISABEL \cite{ISABEL}
intranuclear-cascade models (INC) merged with the 
Dresner evaporation model \cite{Dresner} and the Atchison
fission model (RAL) \cite{RAL},
and using the code INCL by Cugnon {\it et al.} \cite{INCL}
merged in LAHET3 with the ABLA \cite{ABLA} and with
Dresner \cite{Dresner} (+ Atchison \cite{RAL}) evaporation
(+ fission) models, 
with the Dubna transport code CASCADE \cite{CASCADE},
with versions of the Cascade-Exciton Model (CEM)
\cite{CEM} as realized in the codes CEM95 \cite{CEM95} and
CEM2k \cite{CEM2k}, with CEM2k merged \cite{Mashnik02a}-\cite{fitaf}    
with the Generalized Evaporation/fission Model code GEM2 by
Furihata \cite{GEM2}, as well as with the Los Alamos
version of the Quark-Gluon String Model code LAQGSM \cite{LAQGSM}
merged \cite{Mashnik02a}-\cite{fitaf} with  GEM2 \cite{GEM2},
and with versions of the CEM2k and LAQGSM codes both merged 
\cite{Mashnik02a}  with the 
sequential-binary-decay code GEMINI by Charity \cite{GEMINI}.    
Most of these models are widely used to study reaction on heavier 
nuclei (see, {\it e.g.,} \cite{Titarenko02,Titarenko04} and 
references therein).

As we have done previously
(see, {\it e.g.,}~\cite{Titarenko02,Titarenko04}),
we choose here two quantitative criteria to judge how well our
data are described by different models; namely,
the ratio of calculated cross section for the production of a given 
isotope to its measured values
$\sigma^{cal} / \sigma^{exp}$ as a function of the mass number
of products (Fig. 1), and the mean simulated-to-experimental
data ratio (Tables 2 and 3)
\begin{equation}\label{eqf}
\left<F\right>=10^{\sqrt{\left<(\log [ \sigma^{cal} / \sigma^{exp} ] )^2
\right>}}~,
\end{equation}
with its standard deviation~:
\begin{equation}\label{eqsf}
S\left(\left<F\right>\right)=10^{\sqrt{\left<\left(
\left|\log\left(
\sigma^{cal} / \sigma^{exp} \right)
\right|-\log(\left<F\right>)\right)^2\right>}}~.
\end{equation}

For such a comparison, out of all the 76 cross sections measured in this work, 
only 48 were selected to satisfy some rules based on appreciation of the physical principles realized in the models.
For instance, if only an isomer or only the ground state of a nuclide with a 
relatively long-lived isomer was measured, such nuclides were excluded from 
the comparison, but if both were measured separately, their sum was compared 
with calculations. Such rules are essentially similar to those used 
by Titarenko {\it et al.}~\cite{Titarenko02,Titarenko04}.

To understand how different models describe nuclides produced
in the spallation  and fission or fragmentation regions,
we divided all 48 measured nuclides included in the comparison into two 
groups, spallation $(A \ge 95)$ and fission/fragmentation $(A < 95)$.
Table 2 shows values of  $\left<F\right>$ and $S\left(\left<F\right>\right)$
for all compared products (both spallation and fission/fragmentation),
while Tab.\ 3 shows such results only for spallation;
$N$ is the total number of comparisons, $N_{30\%}$---the number of 
comparisons in which
calculated and measured values differ by not more than 30 \%, 
$N_{2.0}$---where the difference was not more than a factor of two.

We note that the codes CEM95 \cite{CEM95} and CEM2k \cite{CEM2k}
consider only competition
between evaporation and fission of excited compound nuclei
and calculate the fission cross sections
for a nuclear reaction on a heavy nucleus, but do not calculate the fission
fragments, as they do not contain a fission model.
The Bertini \cite{Bertini} and ISABEL \cite{ISABEL} INC's are used in our
calculations with the
default options of LAHET3 for evaporation/fission models; they
consider evaporation with the Dresner code \cite{Dresner} and a possible 
fission of heavy compound nuclei using the Atchison RAL fission
model \cite{RAL}, but only if they are heavy enough $(Z > 71)$, 
{\it i.e.,} do not
consider fission for such light targets

\begin{figure}[h!]
\begin{center}
\includegraphics[scale=0.65]{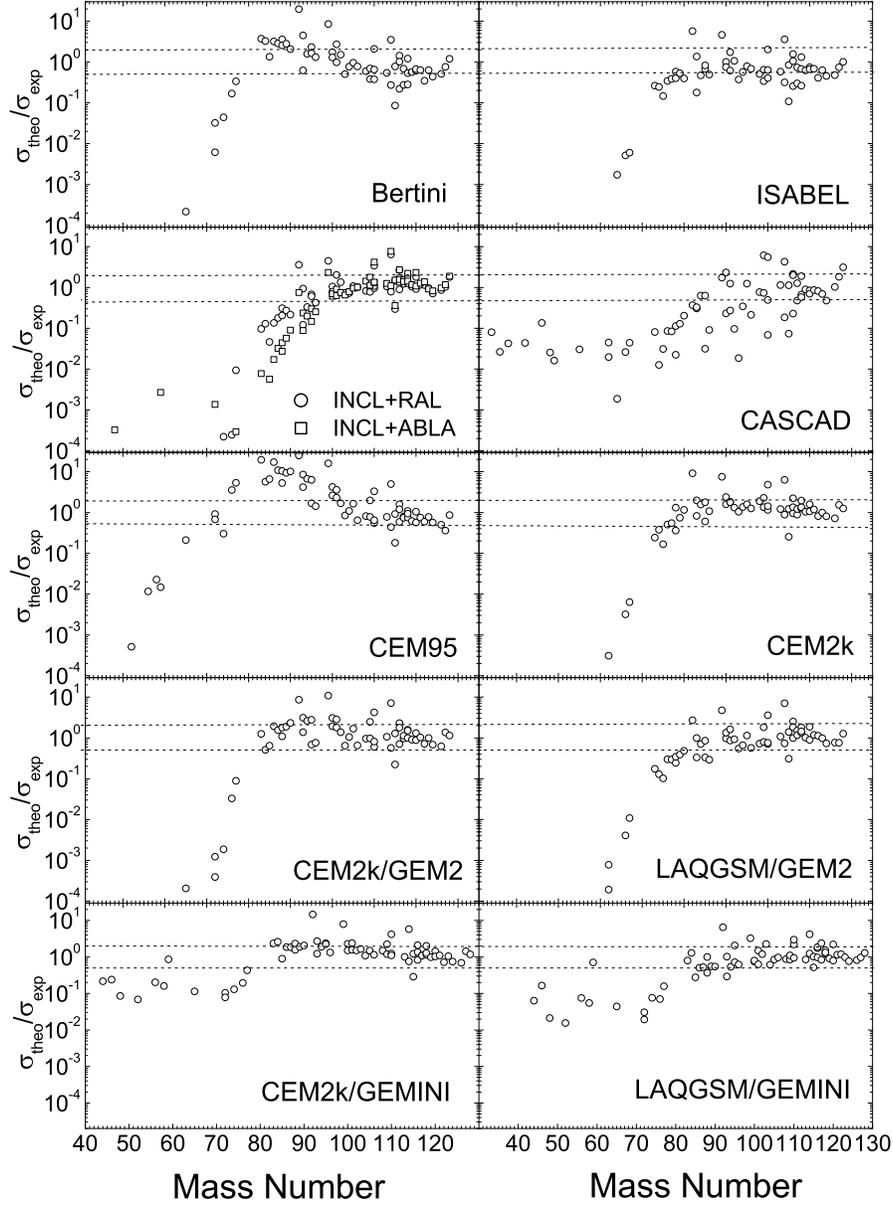}
\caption{\label{fig:fig5f} Comparison of experimental and theoretical
cross sections for our 15\% $^{127}$I + 85\% $^{129}$I target.}
\end{center}
\end{figure}

\begin{table}[h]
  \centering
  \caption{Comparison for all 48 (spallation and fission/fragmentation)
selected isotopes}\label{tab:allA}

\vspace*{3mm}
  \begin{tabular}{lccc}
   Model & $N/N_{30\%}/N_{2.0}$ & $\left<F\right>$ & $S\left(\left<F\right>\right)$ \\
    \hline
    Bertini+Dresner & 43/ 6/23&  6.18& 5.28\\
    ISABEL+Dresner  & 40/ 6/18&  5.70& 4.59\\
    INCL+Dresner    & 36/13/21&  4.10& 3.12\\
    INCL+ABLA       & 36/10/19&  9.95& 6.41\\
    CASCADE         & 48/ 9/16& 11.36& 5.20\\
    CEM95           & 46/ 9/22&  5.15& 3.27\\
    CEM2k           & 40/15/29&  6.50& 6.62\\
    LAQGSM+GEM2     & 42/12/23& 14.37&12.78\\
    CEM2k+GEM2      & 43/11/31& 11.08&10.80\\
    LAQGSM+GEMINI   & 48/17/32&  4.21& 3.35\\
    CEM2k+GEMINI    & 48/12/29&  2.78& 2.11\\
 \end{tabular}
\end{table}

\begin{table}[h]
  \centering
  \caption{Comparison for only 26 selected spallation
isotopes with A $\ge$ 95}\label{tab:Age95}

\vspace*{3mm}
  \begin{tabular}{lccc}
   Model & $N/N_{30\%}/N_{2.0}$ & $\left<F\right>$ & $S\left(\left<F\right>\right)$ \\
    \hline
    Bertini+Dresner & 26/ 6/20 &1.86 &1.47\\
    ISABEL+Dresner  & 26/ 6/17 &1.91 &1.48\\
    INCL+Dresner    & 26/13/21 &2.07 &1.89\\
    INCL+ABLA       & 26/10/19 &2.46 &2.07\\
    CASCADE         & 26/ 9/15 &3.79 &3.02\\
    CEM95           & 26/ 8/20 &1.93 &1.51\\
    CEM2k           & 26/12/22 &1.72 &1.51\\
    LAQGSM+GEM2     & 26/12/21 &1.84 &1.61\\
    CEM2k+GEM2      & 26/ 9/22 &1.84 &1.55\\
    LAQGSM+GEMINI   & 26/16/23 &1.55 &1.40\\
    CEM2k+GEMINI    & 26/ 9/20 &1.75 &1.49\\
  \end{tabular}
\end{table}

{\noindent
as $^{129}$I. 
}
CEM2k+GEM2
and LAQGSM+GEM2 consider fission using the GEM2 model \cite{GEM2}
of only heavy nuclei, with $Z > 65$, {\it i.e.,} also not 
considering fission of our target. Similarly, INCL+ABLA 
\cite{INCL,ABLA} also does not consider fission for I.
Only the code GEMINI by Charity \cite{GEMINI}
merged with CEM2k and LAQGSM considers fission (via  
sequential binary decays) of practically all nuclei, and provides
fission products from our reactions.
This is why CEM2k+GEMINI and LAQGSM+GEMINI agree better than 
all the other models tested here with experimental 
data for this reaction, especially in the A = 40-80 mass region.

Newer calculations \cite{PRC04} have shown that it is possible
to extend the fission model of GEM2 so that it describes also fission
of light nuclei, like $^{129}$I, and gives with CEM2k+GEM2 and LAQGSM+GEM2
for our reactions results very similar 
(even a little better)
to the ones provided by
GEMINI. For this, it is necessary to
fit the ratio of the level-density parameters for the fission
and evaporation channels, $a_f/a_n$. We think that it is possible to extend
in a similar way also the Atchison fission model \cite{RAL} 
and the ABLA evaporation/fission model \cite{ABLA} to describe
fission of Iodine also with the Bertini+Dresner/Atchison, 
ISABEL+Dresner/Atchison,
INCL+Dresner/Atchison, and INCL+ABLA options of LAHET3;
the same is true for the Dubna code CASCADE. Nevertheless,
we are not too optimistic about the predictive power of such
extended versions of these codes as they do not yet contain 
reliable models for fission barriers of light nuclei.

To make the situation even more intricate, we note that when we
merge \cite{OurNewINC} CEM2k+GEM2 and LAQGSM+GEM2 with the Statistical
Multifragmentation Model by Botvina {\it et al.} \cite{SMM}, it is possible to
describe these reactions and get results very similar to the ones
predicted by CEM2k+GEMINI and LAQGSM+GEMINI without extending
the fission model of GEM2, {\it i.e.,} considering only INC, preequilibrium,
evaporation, and multifragmentation processes, but not fission of
$^{129}$I. We will discuss these results in a future publication.
Here, we note that it is impossible to make a correct
choice between fission and fragmentation reaction mechanisms
involved in our $p + ^{129}$I reaction by
comparing theoretical results with only our (or other similar) data;  
addressing this question would require analysis of two- or
multi-particle correlation measurements.

From Fig.\ 1 and Tabs.\ 2 and 3 we see that  
the agreement of different models with our data varies quite a bit.
We find that most of the codes are fairly reliable in predicting 
cross sections for nuclides not too far away in mass from the target, 
but differ greatly in the deep spallation, fission, and/or fragmentation
regions. We conclude that none of the
codes tested here is able to reproduce well all of our data
and all of them need to be further improved; development of a better
universal evaporation/fission model should be of a high priority.

\vspace*{5mm}
{\noindent \bf \large Acknowledgment}\\

We are grateful to the personnel of the JINR phasotron for
providing a high-quality beam. We also thank the Directorate
of the Laboratory of High Energies of JINR for providing us with
radioactive targets.
The work was supported partially by the U.\ S.\ Department of Energy,  the
Moldovan-U.\ S.\ Bilateral Grants Program, CRDF Projects MP2-3025
and MP2-3045, and by the NASA ATP01 Grant NRA-01-01-ATP-066.

\end{document}